\documentclass[10pt]{article}

\usepackage{amsxtra,amssymb,amsthm,amsmath,latexsym,amsfonts}

\usepackage{graphicx}

\newcommand{\bee}{\begin{equation*}}
\newcommand{\eee}{\end{equation*}}
\newcommand{\be}{\begin{equation}}
\newcommand{\ee}{\end{equation}}

\title{Heat transfer in a complex medium
}
\author{A. G. Ramm\\
\small Department of Mathematics\\[-0.8ex]
\small Kansas State University, Manhattan, KS 66506-2602, USA\\[-0.8ex]
\small \texttt{ramm@math.ksu.edu}\\
}
\begin{document}


\date{} \maketitle

\begin{abstract} 
The heat equation is considered in
the complex medium consisting of many small bodies (particles) embedded
in a given material. On the surfaces of the small bodies an impedance
boundary condition is imposed. An equation for the limiting field is
derived when the characteristic size $a$ of the small bodies tends to
zero, their total number $\mathcal{N}(a)$ tends to infinity at a suitable
rate, and the distance $d = d(a)$ between neighboring small bodies tends
to zero: $a << d$, $\lim_{a\to 0}\frac{a}{d(a)}=0$. No periodicity is assumed about the distribution of the
small bodies. These results are basic for a method of creating a medium
in which heat signals are transmitted along a given line. The technical
part for this method is based on an inverse problem of finding potential with
prescribed eigenvalues.

 \end{abstract}

Keywords:

heat transfer; many-body problem; transmission of heat signals; inverse problems; materials science.

MSC 80M40; 80A20, 35B99; 35K20; 35Q41; 35R30;74A30; 74G75

PACS 65.80.-g

\section{Introduction and results} \label{Introduction and results}

In this paper the problem of heat transfer in a complex medium
consisting of many small impedance particles of an arbitrary shape is
solved. Equation for the effective limiting temperature is derived
when the characteristic size $a$ of the particles tends to zero while
their number tends to infinity at a suitable rate while the distance $d$ between
closest neighboring particles is much larger than $a$, $d>>a$.

These results are used for developing a method for creating materials
in which heat is transmitted along a line. Thus, the information can
be transmitted by a heat signals.

The contents of this paper is based on the earlier papers of the author
cited in the bibliography, especially  \cite{R635}, \cite{R624} and \cite{R655}.

Let many small
bodies (particles) $D_m$, $1 \leq m \leq M$, of an arbitrary shape
 be distributed in a
bounded domain $D \subset \mathbb{R}^3$, diam$D_m =
2a$, and the boundary of $D_m$ is denoted by $\mathcal{S}_m$ and is
assumed twice continuously differentiable.
The small bodies are distributed according to the law
\begin{equation}\label{eq:1} \mathcal{N}(\Delta) = \frac{1}{a^{2 -
\kappa}}\int_{\Delta} N(x)dx[1 + o(1)] ,\quad a \rightarrow 0. \end{equation}
Here $\Delta \subset D$ is an arbitrary open subdomain of
$D$, $\kappa \in [0, 1)$ is a constant, $N(x) \geq 0$ is a
continuous function, and $\mathcal{N}(\Delta)$ is the number of the small
bodies $D_m$ in $\Delta$. The heat equation can be stated as
follows: \begin{equation}\label{eq:2} u_t = \nabla^2u + f(x) \, \textrm{
in } \, \mathbb{R}^3 \setminus \displaystyle\bigcup^{M}_{m =1}  D_m,:=\Omega,
\quad u|_{t=0}=0,
\end{equation}
\begin{equation}\label{eq:3} u_N = \zeta_m u \textrm{ on } \mathcal{S}_m,\quad
1 \leq m \leq M, \qquad Re \zeta_m\ge 0. \end{equation}
Here $N$ is the outer unit normal to
$\mathcal{S}$,
$$\mathcal{S}:= \displaystyle\bigcup^{M}_{m = 1} \mathcal{S}_m,\quad \zeta_m
= \displaystyle\frac{h(x_m)}{a^{\kappa}}, \quad x_m \in D_m,\quad 1 \leq m
\leq M,$$  and $h(x)$ is a continuous function in $D$, Re$h\ge 0$.\\
Denote
$$\mathcal{U} := \mathcal{U}(x, \lambda) =
\displaystyle\int^{\infty}_{0}e^{-\lambda t}u(x, t)dt.$$
Then, taking the Laplace transform of equations (\ref{eq:2})
- (\ref{eq:3}) one gets:
\begin{equation}\label{eq:4} -\nabla^2\mathcal{U} +
\lambda\mathcal{U} = \lambda^{-1}f(x) \textrm{ in } \Omega, \end{equation}
\begin{equation}\label{eq:5} \mathcal{U}_N = \zeta_m\mathcal{U} \textrm{
on } \mathcal{S}_m, 1 \leq m \leq M. \end{equation}
Let
\begin{equation}\label{eq:6} g(x,y) := g(x, y, \lambda):=
\frac{e^{-\sqrt{\lambda}|x - y|}}{4\pi|x - y|},
\end{equation}
\begin{equation}\label{eq:6'}
F(x, \lambda):=\frac{1}{\lambda}\int_{\mathbb{R}^3}g(x,y)f(y)dy.
\end{equation} Look for the solution to (\ref{eq:4}) - (\ref{eq:5}) of the
form \begin{equation}\label{eq:7} \mathcal{U}(x, \lambda) = F(x, \lambda)
+ \sum^{M}_{m = 1}\int_{\mathcal{S}_m}g(x, s)\sigma_m(s)ds,
\end{equation}
where
\begin{equation}\label{eq:6''}
\mathcal{U}(x, \lambda):=\mathcal{U}(x):=\mathcal{U},
\end{equation}
and $\mathcal{U}(x)$ depends on $\lambda$.

The functions $\sigma_m$ are unknown and should be found from the boundary
conditions (\ref{eq:5}). Equation (\ref{eq:4}) is satisfied by
$\mathcal{U}$ of the form (\ref{eq:7}) with arbitrary continuous $\sigma_m$. To satisfy
the boundary condition
(\ref{eq:5}) one has to solve the following equation obtained from the boundary condition   (\ref{eq:5}):
\begin{equation}\label{eq:8}
\frac{\partial \mathcal{U}_e(x)}{\partial N} + \frac{A_m\sigma_m -
\sigma_m}{2} - \zeta_m\mathcal{U}_e - \zeta_mT_m\sigma_m = 0 \textrm{ on }
\mathcal{S}_m,\quad 1 \leq m \leq M, \end{equation}
where the effective field $ \mathcal{U}_e(x)$ is defined by the formula:
\begin{equation}\label{eq:9} \mathcal{U}_e(x) := \mathcal{U}_{e,m}(x) :=
\mathcal{U}(x) - \int_{\mathcal{S}_m}g(x, s)\sigma_m(s)ds,
\end{equation}
the operator $T_m$ is defined by the formula:
\begin{equation}\label{eq:10} T_m\sigma_m = \int_{\mathcal{S}_m}g(s,
s')\sigma_m(s')ds',
 \end{equation}
 and $A_m$ is:
 \begin{equation}\label{eq:9'}
  A_m\sigma_m = 2\int_{\mathcal{S}_m}\frac{\partial
g(s, s')}{\partial N_s}\sigma_m(s')ds'. \end{equation}
 In deriving equation (\ref{eq:8}) we have used the known
formula for the outer limiting value on $\mathcal{S}_m$ of the normal
derivative of a simple layer potential.

We now apply the ideas
and methods for solving many-body scattering problems developed in
\cite{R635} - \cite{R633}.$\\\\$ 
Let us call $\mathcal{U}_{e,m}$ the
effective (self-consistent) value of $\mathcal{U}$, acting on the $m$-th body.
As $a \rightarrow 0$, the dependence on $m$ disappears, since
$$\displaystyle\int_{\mathcal{S}_m}g(x, s)\sigma_m(s)ds \rightarrow 0
\textrm{ as } a \rightarrow 0.$$
One has
\begin{equation}\label{eq:11}
\mathcal{U}(x, \lambda) = F(x, \lambda) + \sum^{M}_{m = 1}g(x, x_m)Q_m +
\mathcal{J}_2, \quad x_m \in D_m, \end{equation} where
$$Q_m :=\int_{\mathcal{S}_m}\sigma_m(s)ds,$$
\begin{equation}\label{eq:12}
\mathcal{J}_2 := \sum^{M}_{m = 1}\int_{\mathcal{S}_m}[g(x, s') - g(x,
x_m)]\sigma_m(s')ds'.
\end{equation}
Define
 \begin{equation}\label{eq:12'}
 \mathcal{J}_1 := \sum^{M}_{m = 1}g(x, x_m)Q_m.
\end{equation}
We prove in Lemma 3, Section 4 (see also \cite{R635} and \cite{R624}) that
\begin{equation}\label{eq:13}
|\mathcal{J}_2| << |\mathcal{J}_1| \textrm{ as } a \rightarrow 0
\end{equation} provided that \begin{equation}\label{eq:14} \lim_{a
\rightarrow 0}\frac{a}{d(a)} = 0, \end{equation}
where $d(a)=d$ is the minimal distance between neighboring particles.

 If
(\ref{eq:13}) holds,
then problem (\ref{eq:4}) - (\ref{eq:5}) is solved asymptotically by the
formula \begin{equation}\label{eq:15} \mathcal{U}(x, \lambda) = F(x,
\lambda) + \sum^{M}_{m = 1}g(x, x_m)Q_m, \quad a \rightarrow 0,
\end{equation} provided that asymptotic formulas for $Q_m$, as $a
\rightarrow 0$, are found.

To find formulas for $Q_m$, let us
integrate (\ref{eq:8}) over $\mathcal{S}_m$, estimate the order of the
terms in the resulting equation as $a \rightarrow 0$, and keep the main terms,
that is, neglect the terms of higher order of smallness as $a\to 0$.

 We get
\begin{equation}\label{eq:16} \int_{\mathcal{S}_m}\frac{\partial
\mathcal{U}_e}{\partial N}ds = \int_{D_m}\nabla^2\mathcal{U}_edx
= O(a^3). \end{equation}
Here we assumed that $|\nabla^2\mathcal{U}_e| =
O(1), a \rightarrow 0$. This assumption is valid since
$\mathcal{U}=\lim_{a\to 0} \mathcal{U}_e$ is smooth
as a solution to an elliptic equation. One has
\begin{equation}\label{eq:17} \int_{\mathcal{S}_m}\frac{A_m\sigma_m -
\sigma_m}{2}ds = - Q_m[1 + o(1)], \, a \rightarrow 0. \end{equation}
This
relation is proved in Lemma 2, Section 4, see also \cite{R635}. Furthermore,
\begin{equation}\label{eq:18} -\zeta_m\int_{\mathcal{S}_m}\mathcal{U}_eds
= -\zeta_m|\mathcal{S}_m|\mathcal{U}_e(x_m) = O(a^{2 - \kappa}), \quad a
\rightarrow 0, \end{equation} where $|\mathcal{S}_m| = O(a^2)$ is the
surface area of $\mathcal{S}_m$. Finally,
$$-\zeta_m\int_{\mathcal{S}_m}ds\int_{\mathcal{S}_m}g(s,
s')\sigma_m(s')ds' =
-\zeta_m\int_{\mathcal{S}_m}ds'\sigma_m(s')\int_{\mathcal{S}_m}dsg(s,
s')$$ \begin{equation}\label{eq:19}
 = Q_m O(a^{1 - \kappa}), \qquad a \rightarrow 0. \end{equation}
 Thus, the
main term of the asymptotic of $Q_m$, as $a\to 0$, is
\begin{equation}\label{eq:20} Q_m
= -\zeta_m|\mathcal{S}_m|\mathcal{U}_e(x_m). \end{equation} Formulas
(\ref{eq:20}) and (\ref{eq:15}) yield
\begin{equation}\label{eq:21}
\mathcal{U}(x, \lambda) = F(x, \lambda) - \sum^{M}_{m =
1}g(x,x_m)\zeta_m|\mathcal{S}_m|\mathcal{U}_e(x_m, \lambda),
\end{equation} and
\begin{equation}\label{eq:22} \mathcal{U}_e(x_m, \lambda) = F(x_m,
\lambda) - \sum^{M}_{m' \neq m, m' = 1}g(x_m,
x_{m'})\zeta_{m'}|\mathcal{S}_{m'}|\mathcal{U}_e(x_{m'}, \lambda).
\end{equation}
 Denote $$\mathcal{U}_e(x_m, \lambda) := \mathcal{U}_m,\quad
F(x_m, \lambda) := F_m,\quad g(x_m, x_{m'}) := g_{mm'},$$
 and write (\ref{eq:22}) as a linear algebraic system 
 for $\mathcal{U}_m$:
 \begin{equation}\label{eq:23}
\mathcal{U}_m = F_m - a^{2 - \kappa}\sum_{m' \neq
m}g_{mm'}h_{m'}c_{m'}\mathcal{U}_{m'},\quad 1 \leq m \leq M,
\end{equation}
where $h_{m'} = h(x_{m'})$,  $\zeta_{m'} = \frac{h_{m'}}{a^{\kappa}}$, $c_{m'}
:= |S_{m'}|a^{-2}$. $\\$ Consider a partition of the bounded domain
$D$, in which the small bodies are distributed, into a union of
$P << M$ small nonintersecting cubes $\Delta_p$, $1 \leq p \leq P$, of
side $b$,
 $$b >> d, \quad b = b(a) \rightarrow 0\quad \text{as}\,\,  a \rightarrow 0\quad \lim_{a\to 0}\frac{d(a)}{b(a)}=0.$$
 Let $x_p \in
\Delta_p$, $|\Delta_p| =$ volume of $\Delta_p$. One has $$a^{2 -
\kappa}\sum^{M}_{m' = 1, m' \neq m}g_{mm'}h_{m'}c_{m'}\mathcal{U}_{m'} =
a^{2 - \kappa}\sum^{P}_{p' = 1, p' \neq
p}g_{pp'}h_{p'}c_{p'}\mathcal{U}_{p'}\sum_{x_{m'} \in \Delta_{p'}}1 =$$
\begin{equation}\label{eq:24} = \sum_{p' \neq
p}g_{pp'}h_{p'}c_{p'}\mathcal{U}_{p'}N(x_{p'})|\Delta_{p'}|[1 + o(1)],
\quad a \rightarrow 0. \end{equation}
Thus, (\ref{eq:23}) yields a linear algebraic system (LAS) of order $P<<M$
for the unknowns $ \mathcal{U}_p$:
\begin{equation}\label{eq:25} \mathcal{U}_p = F_p - \sum^{P}_{p' \neq p,
p' = 1}g_{pp'}h_{p'}c_{p'}N_{p'}\mathcal{U}_{p'}|\Delta_{p'}|, \quad 1
\leq p
\leq P. \end{equation}
Since $P<<M$, the order of the original LAS (\ref{eq:23}) is drastically reduced.
This is crucial when the number of particles tends to infinity and their size $a$
tends to zero.
We have assumed that \begin{equation}\label{eq:26}
h_{m'} = h_{p'}[1 + o(1)],\quad c_{m'} = c_{p'}[1 + o(1)],\quad \mathcal{U}_{m'} =
\mathcal{U}_{p'}[1 + o(1)], \,\,a \rightarrow 0, \end{equation} for
$x_{m'}
\in \Delta_{p'}$. This assumption is justified, for example,  if the functions $h(x)$,
$\mathcal{U}(x, \lambda)$,
$$c(x) = \displaystyle\lim_{x_{m'} \in \Delta_x,
a \rightarrow 0}\displaystyle\frac{|S_{m'}|}{a^2},$$
 and $N(x)$ are
continuous, but these assumptions can be relaxed.

The continuity of the $\mathcal{U}(x, \lambda)$ is a consequence of the
fact that this function satisfies elliptic equation,
 and the continuity of $c(x)$ is assumed. If all the small
bodies are identical, then $c(x) = c = const$,
so in this case the function $c(x)$ is certainly continuous.

The sum in the
right-hand side of (\ref{eq:25}) is the Riemannian sum for the integral
\newpage
\begin{equation}  lim_{a \rightarrow 0}\sum^{P}_{p' = 1, p'
\neq p}g_{pp'}h_{p'}c_{p'}N(x_{p'})\mathcal{U}_{p'}|\Delta_p'|= \nonumber
\end{equation}
\begin{equation}\int_{D}g(x,y)h(y)c(y)N(y)\mathcal{U}(y, \lambda)dy\nonumber
\end{equation}

Therefore, linear algebraic system (\ref{eq:25}) is a
collocation method for solving integral equation
\begin{equation}\label{eq:28} \mathcal{U}(x, \lambda) = F(x, \lambda) -
\int_{D}g(x,y) c(y) h(y) N(y)\mathcal{U}(y, \lambda)dy.
\end{equation} Convergence of this method for solving equations with
weakly singular kernels is proved in \cite{R563},
see also \cite{R573} and \cite{R599}.

 Applying the
operator $-\nabla^2 + \lambda$ to equation (\ref{eq:28})
one gets an elliptic differential equation:
\begin{equation}\label{eq:28'}
(-\Delta +\lambda)\mathcal{U}(x, \lambda)=\frac{f(x)}{\lambda} -c(x) h(x) N(x)\mathcal{U}(x, \lambda).
\end{equation}
 Taking the inverse Laplace
transform of this equation yields
\begin{equation}\label{eq:29} u_t =
\Delta u + f(x) - q(x)u, \quad q(x) :=c(x) h(x) N(x). \end{equation}
 Therefore,
 the limiting equation for the temperature contains the term
$q(x)u$. Thus, the embedding of many small particles creates a
distribution of source and sink terms in the medium, the distribution of
which is described by the term $q(x)u$.$\\$ If one solves equation
(\ref{eq:28}) for $\mathcal{U}(x, \lambda)$, or linear algebraic system
(\ref{eq:25}) for $\mathcal{U}_p(\lambda)$, then one can Laplace-invert
$\mathcal{U}(x, \lambda)$ for $\mathcal{U}(x, t)$. Numerical methods for
Laplace inversion from the real axis are discussed in \cite{R198} -
\cite{R569}.$\\$ If one is interested only in the average temperature, one
can use the relation \begin{equation}\label{eq:30} \lim_{T \rightarrow
\infty}\frac{1}{T}\int^{T}_{0}u(x, t)dt = \lim_{\lambda \rightarrow
0}\lambda\mathcal{U}(x, \lambda). \end{equation}
Relation
(\ref{eq:30}) is proved in Lemma 1, Section 4. It holds if the limit on one of
its sides exists. The limit on the right-hand side of (\ref{eq:30}) let us
denote by $\psi(x)$.
From equations  (7) and  (31) it follows that $\psi$ satisfies the equation
$$\psi=\varphi- B\varphi,$$
where
$$\varphi:= \int_{\Omega}g_0(x,y)f(y)dy,$$
$$g_0(x,y):=\frac{1}{4\pi |x-y|},$$
$$B\psi:= \int_{\Omega}g_0(x,y)q(y)\psi(y)dy,$$
and
$$  q(x):=c(x)h(x)N(x).$$
The function $\psi$  can be
calculated by the formula
\begin{equation}\label{eq:31} \psi(x) = (I +
B)^{-1}\varphi.
 \end{equation}
  From the physical point of view the function $h(x)$ is
non-negative because the flux $-\nabla u$ of the heat flow is proportional
to the temperature $u$ and is directed along the outer normal $N$: $-u_N =
h_1u$, where $h_1 = -h < 0$. Thus, $q \ge
 0$.

 It is proved in \cite{R203}
- \cite{R227} that zero is not an eigenvalue of the operator $-\nabla^2 +
q(x)$ provided that $q(x) \geq 0$ and
$$q = O\big(\displaystyle\frac{1}{|x|^{2 + \epsilon}} \big), \quad |x|\to \infty, $$
and $ \epsilon > 0.$

In our case, $q(x) = 0$ outside of the bounded region
$D$, so the operator $(I + B)^{-1}$ exists and is bounded
in $C(D)$.

Let us formulate our basic result.

{\bf Theorem 1.} {\it Assume (\ref{eq:1}),
(\ref{eq:14}), and $h \ge 0$. Then, there exists the limit
$\mathcal{U}(x, \lambda)$ of $\mathcal{U}_e(x, \lambda)$ as $a
\rightarrow
0$, $\mathcal{U}(x, \lambda)$ solves equation (\ref{eq:28}), and there
exists the limit (\ref{eq:30}), where $\psi(x)$ is given by formula
(\ref{eq:31}).}

Methods of our proof of Theorem 1 are quite different from the
proof of homogenization theory results in \cite{JKO} and \cite{MK}.

The author's plenary talk at Chaos-2015 Conference was published in \cite{R652}.

\section{Creating materials which allows one\\ to transmit heat signals along a line} \label{Creating materials which allows one to transmit heat signals along a line}

In applications it is of interest to have materials in which heat propagates along a line and decays fast in all the directions orthogonal to this line.\\

In this Section a construction of such material is given. We follow \cite{R655} with some simplifications.

 The idea is to create first the medium in which the heat transfer is governed by the equation
\begin{equation}\label{eq1}
u_t = \Delta u - q(x)u \quad \text{in } D, \quad u|_S = 0, \quad u|_{t=0}=f(x),
\end{equation}
where $D$ is a bounded domain with a piece-wise smooth boundary $S$, $D = D_0 \times [0, L]$, $D_0 \subset \mathbb{R}^2$ is  a smooth domain orthogonal to the axis $x_1$, $x = (x_1, x_2, x_3)$, $x_2, x_3 \in D_0$, $0 \leq x_1 \leq L$.\\

Such a medium is created by embedding many small impedance particles into a given domain $D$ filled with a homogeneous material. A detailed argument, given in Section 1 (see also \cite{R635} and \cite{R624}), yields the following result.\\

Assume that in every open subset $\Delta$ of $D$ the number of small particles is defined by the formula:
\begin{equation}\label{eq2}
\mathcal{N}(\Delta) = \frac{1}{a^{2 - \kappa}}\int_{\Delta}N(x)dx[1 + o(1)], \quad a \to 0,
\end{equation}
where $a > 0$ is the characteristic size of a small particle, $\kappa \in [0,1)$ is a given number and $N(x) \geq 0$ is a continuous in $D$ function.\\

Assume also that on the surface $S_m$ of the $m$-th particle $D_m$ the impedance boundary condition holds. Here
 $$1 \leq m \leq M = \mathcal{N}(D) = O\left( \frac{1}{a^{2 - \kappa}} \right), \quad a\to 0,$$
 and the impedance boundary conditions are:
\begin{equation}\label{eq3}
u_N = \zeta_m u \quad \text{on } S_m, \quad \text{Re}\zeta_m \geq 0,
\end{equation}
where
$$\zeta_m := \frac{h(x_m)}{a^\kappa}$$
is the boundary impedance, $x_m \in D_m$ is an arbitrary point (since $D_m$ is small
the position of $x_m$ in $D_m$ is not important), $\kappa$ is the same parameter as in \eqref{eq2} and $h(x)$ is a continuous in $D$ function, Re$h \geq 0$, $N$ is the unit normal to $S_m$ pointing out of $D_m$. The functions $h(x)$, $N(x)$ and the number $\kappa$ can be chosen
as the experimenter wishes.\\

It is proved in Section 1 (see also \cite{R635}, \cite{R624}) that, as $a \to 0$, the solution of the problem
\begin{equation}\label{eq4}
u_t = \Delta u \quad \text{in } D \setminus \displaystyle\bigcup_{m = 1}^M D_m, \, u_N = \zeta_m u \quad \text{on } S_m, \, 1 \leq m \leq M, \,
\end{equation}
\begin{equation}\label{eq4'}
u|_S = 0,
\end{equation}
and
\begin{equation}\label{eq5}
u|_{t = 0} = f(x),
\end{equation}
has a limit $u(x,t)$. This limit  solves problem \eqref{eq1} with
\begin{equation}\label{eq6}
q(x) = c_S N(x) h(x),
\end{equation}
where
\begin{equation}\label{eq6'}
\quad c_S := \frac{|S_m|}{a^2}=const,
\end{equation}
and $|S_m|$ is the surface area of $S_m$. By assuming that $c_S$ is a constant, we assume, for simplicity only,
 that the small particles are identical in shape, see \cite{R635}.

 Since $N(x)\ge 0$ is an arbitrary
continuous function and $h(x)$, $Re h\ge 0$, is an arbitrary continuous function,
and both functions can be chosen by experimenter as he/she wishes, it is clear
that an arbitrary real-valued potential $q$ can be obtained by formula \eqref{eq6}.\\

Suppose that
\begin{equation}\label{eq7}
(-\Delta + q(x))\phi(x) = \lambda_n\phi_n, \quad \phi_n|_S = 0, \quad ||\phi_n||_{L^2(D)} = ||\phi_n|| = 1,
\end{equation}
where $\{\phi_n\}$ is an orthonormal basis of $L^2(D) := H$. Then the unique solution to \eqref{eq1} is
\begin{equation}\label{eq8}
u(x, t) = \sum_{n = 1}^\infty e^{-\lambda_n t}(f, \phi_n)\phi_n(x).
\end{equation}
If $q(x)$ is such that $\lambda_1 = 0$, $\lambda_2 \gg 1$, and $\lambda_2 \leq \lambda_3 \leq \dots$,  then, as $t \to \infty$, the series \eqref{eq8} is well approximated by its first term
\begin{equation}\label{eq9}
u(x, t) = (f, \phi_1)\phi_1 + O(e^{-10t}), \quad t \to \infty.
\end{equation}
If $\lambda_1>0$ is very small, then the main term of the solution is
$$u(x, t) = (f, \phi_1)\phi_1e^{-\lambda_1 t} + O(e^{-10t})$$
as
$ t \to \infty$. The term $e^{-\lambda_1 t}\sim 1$ if $t<<\frac{1}{\lambda_1}$.

Thus, our problem is solved if $q(x)$ has the following property:
\begin{equation}\label{eq10}
|\phi_1(x)| \text{ decays as $\rho$ grows}, \quad \rho = (x_2^2 + x_3^2)^{1/2}.
\end{equation}
Since the eigenfunction is normalized, $||\phi_1||=1$, this function will not tend to zero
in a neighborhood of the line $\rho=0$, so information can be transformed by the heat signals along the line
$\rho=0$, that is, along $s-$axis. Here we use the cylindrical coordinates:
$$x=(x_1,x_2,x_3)=(s, \rho, \theta), \quad
s=x_1, \quad \rho= (x_2^2+x_3^2)^{1/2}.$$
 In Section 3 the domain $D_0$ is a disc and the potential $q(x)$ does not depend on $\theta$.

The technical part of solving our problem consists of the construction of $q(x) = c_S N(x) h(x)$ such that
\begin{equation}\label{eq11}
\lambda_1 = 0, \quad \lambda_2 \gg 1; \quad |\phi_1(x)| \text{ decays as $\rho$ grows}.
\end{equation}
Since the function $N(x) \geq 0$ and $h(x), \text{Re}h \geq 0,$ are at our disposal, any desirable $q, \text{Re}\,q \geq 0$,
can be obtained by embedding many small impedance particles in a given domain $D$.
In Section 3, a potential $q$ with the desired properties is constructed.
 This construction allows one to transform information along a straight line using heat signals.

\section{Construction of $q(x)$}\label{Section2}

Let $$q(x) = p(\rho)+Q(s),$$ where $s := x_1$, $\rho:=(x_2^2+x_3^2)^{1/2}$.
Then the solution to problem \eqref{eq7} is $u = v(\rho)w(s)$, where
\begin{multline}\label{eq12}
-v_m'' - \rho^{-1} v_m' + p(\rho)v_m = \mu_mv_m, \quad 0 \leq \rho \leq R, \\
|v_m(0)| < \infty, \quad v_m(R) = 0,
\end{multline}
and
\begin{multline}\label{eq12'}
 -w_l'' + Q(s)w_l = \nu_l w_l, \quad 0 \leq s \leq L,\\
 w_l(0) = 0, \quad w_l(L) = 0.
\end{multline}
One has
\begin{equation}\label{eq13}
 \lambda_n=\mu_m + \nu_l, \quad n = n(m,l).
\end{equation}

Our task is to find a potential $Q(s)$ such that $\nu_1 = 0$, $\nu_2 \gg  1$ and a potential $p(\rho)$ such that $\mu_1 = 0, \mu_2 \gg 1$ and $|v_m(\rho)|$ decays as $\rho$ grows.\\

It is known how to construct $q(s)$ with the desired properties: the Gel'fand-Levitan  method allows one to do this, see
\cite{R470}. Let us recall this construction. One has $\nu_{l0} = l^2$, where we set $L = \pi$ and denote by
 $\nu_{l0}$ the eigenvalues of the problem  \eqref{eq12'} with $Q(s)=0$.
 Let the eigenvalues of the operator  \eqref{eq12'} with $Q\neq 0$ be
  $\nu_1 = 0, \nu_2 = 11, \nu_3 = 14,  \nu_l = \nu_{l0}$ for $l \ge 4$.\\

The kernel $L(x,y)$ in the Gel'fand-Levitan theory is defined as follows:
$$
L(x,y)=\int_{-\infty}^{\infty} \frac{\sin(\sqrt{\lambda}x)}{\sqrt{\lambda}} \frac{\sin(\sqrt{\lambda}y)}{\sqrt{\lambda}}d(\varrho (\lambda)-\varrho_0(\lambda)),
$$
where $\varrho(\lambda)$ is the spectral function of the operator \eqref{eq12'} with the potential $Q=Q(s)$,
and  $\varrho_0(\lambda)$ is the spectral function of the operator \eqref{eq12'} with the potential $Q=0$ and the same
boundary conditions as for the operator with $Q\neq 0$.

Due to our choice of $\nu_l$ and the normalizing constants $\alpha_j$, namely: $\alpha_j=\frac{\pi}{2}$ for $j\ge 2$
and $\alpha_1=\frac{\pi^3}{3}$,  the kernel $L(x,y)$ is given explicitly by the formula:
\begin{multline}\label{eq14}
L(x,y) =\frac{3xy}{\pi^3}+ \frac{2}{\pi}\Big(\frac{\sin (\sqrt{\nu_2} x)}{\sqrt{\nu_2}} \frac{\sin (\sqrt{\nu_2} y)}{\sqrt{\nu_2}} +\frac{\sin (\sqrt{\nu_3} x)}{\sqrt{\nu_3}} \frac{\sin (\sqrt{\nu_3} y)}{\sqrt{\nu_3}} \Big) - \\
  -\frac{2}{\pi}\Big(\sin x \sin  y +\sin (2 x) \sin (2 y)  +\sin (3 x) \sin (3 y)\Big),
\end{multline}
where $\nu_1=0$, $\nu_2=11$ and $\nu_3=14$.
This is a finite rank kernel. The term $xy$
is the value of the function $\frac{\sin \nu x}{\nu} \frac{\sin \nu y}{\nu}$ at $\nu=0$, and the
corresponding normalizing constant is $\frac{\pi^3}{3}=||x||^2=\int_0^{\pi}x^2dx$.

Solve the Gel'fand-Levitan  equation:
\begin{equation}\label{eq15}
K(s, \tau) + \int_0^sK(s,s')L(s', \tau)ds' = -L(s, \tau), \quad 0 \leq \tau \leq s,
\end{equation}
which is uniquely solvable (see  \cite{R470}). Since equation \eqref{eq15} has finite-rank kernel it can be
solved analytically being equivalent to a linear algebraic system.

If the function $K(s, \tau)$ is found, then
the potential $Q(s)$ is computed by the formula (\cite{L}, \cite{R470}):
\begin{equation}\label{eq16}
Q(s) = 2\frac{dK(s,s)}{ds},
\end{equation}
and this $Q(s)$ has the required properties: $\nu_1 = 0, \nu_2 \gg 1, \nu_l \leq \nu_{l + 1}$.

Consider now the operator \eqref{eq12} for $v(\rho)$. Our problem is to calculate
 $p(\rho)$ which has the required properties:
 $$\mu_1 = 0, \quad \mu_2 \gg 1,\quad \mu_m \leq \mu_{m + 1},$$
 and $ |\phi_m(\rho)|$ decays as $\rho$ grows.\\

We reduce this problem to the previous one that was solved above. To do this, set $v = \frac{\psi}{\sqrt{\rho}}$.
 Then equation $$-v'' - \frac{1}{\rho}v' + p(\rho)v = \mu v,$$ is transformed to the equation
\begin{equation}\label{eq17}
-\psi'' - \frac{1}{4\rho^2}\psi + p(\rho)\psi = \mu\psi.
\end{equation}

Let
\begin{equation}\label{eq17'}
p(\rho) = \frac{1}{4\rho^2} + Q(\rho),
\end{equation}
where $Q(\rho)$ is constructed above. Then equation \eqref{eq17} becomes
\begin{equation}\label{eq18}
-\psi'' + Q(\rho)\psi = \mu\psi,
\end{equation}
and the boundary conditions are:
\begin{equation}\label{eq18'}
 \psi(R) = 0, \quad \psi(0)=0.
\end{equation}
The problem  \eqref{eq18}- \eqref{eq18'} has the desired eigenvalues $\mu_1 = 0, \mu_2 \gg 1, \mu_m \leq \mu_{m + 1}$.

 The eigenfunction $$\phi_1(x) = v_1(\rho)w_1(s),$$ where $v_1(\rho) = \frac{\psi_1(\rho)}{\sqrt{\rho}}$,
 decays as $\rho$ grows, and the eigenvalues $\lambda_n$ can be calculated by the formula:
  $$\lambda_n = \mu_m + \nu_l, \quad m,l\ge 1, \quad n=n(m,l).$$
 Since $\mu_1 = \nu_1 = 0$ one has $\lambda_1 = 0$. Since $\nu_2 = 11$ and $ \mu_2 = 11$, one has  $\lambda_2 = 11 \gg 1$.

 Thus, the desired potential is constructed:
 $$q(x) = Q(s)+(\frac{1}{4\rho^2} + Q(\rho)),$$ where $Q(s)$ is given by formula \eqref{eq16}.\\

This concludes the description of our procedure for the construction of $q$.\\

\textbf{Remark 1.} It is known (see, for example, \cite{L}) that the normalizing constants
 $$\alpha_j := \int_0^\pi\varphi_j^2(s)ds$$ and the eigenvalues $\lambda_j$,
  defined by the differential equation
   $$-\frac {d^2\varphi_j}{ds^2} + Q(s)\varphi_j = \lambda_j\varphi_j,$$
   the boundary conditions
$$   \varphi_j'(0) = 0,\quad \varphi_j'(\pi) = 0,$$
 and the normalizing condition $\varphi_j(0) = 1$,  have the following asymptotic:
$$\alpha_j = \frac{\pi}{2} + O(\frac{1}{j^2}) \quad \text{as } j \to \infty,$$
and $$ \sqrt{\lambda_j} =
 j + O(\frac{1}{j}) \quad \text{as } j \to \infty.$$
 The differential equation
 $$-\Psi_j^{''}+Q(s)\Psi_j=\nu_j\Psi_j,$$
 the boundary condition
 $$\Psi_j(0)=0, \quad \Psi_j(\pi)=0,$$
 and the normalizing condition $\Psi_j'(0)=1$
 imply
 $$ \sqrt{\lambda_j} =j + O(\frac{1}{j}) \quad \text{as } \quad j \to \infty,$$
 $$\Psi_j(s)\sim \frac{\sin (js)}{j}\quad \text {as}\quad j\to \infty.$$
  The main term of the normalized eigenfunction is:
 $$\frac {\Psi_j}{||\Psi_j||}\sim \sqrt{2/\pi}\sin (js)  \quad\text {as}  \quad  j\to \infty, $$
 and the main term of the normalizing constant is:
 $$\alpha_j\sim \frac{\pi}{2j^2}   \quad\text {as}  \quad  j\to \infty.$$

\section{Auxiliary results}\label{Auxiliary results}

\textbf{Lemma 1} {\em If one of the limits $\lim_{t\to \infty}\frac{1}{t}\int^{t}_{0}u(s)ds$ or
$\lim_{\lambda\to 0}\lambda \mathcal{U}(\lambda)$ exists, then the other also exists and
they are equal to each other:
$$\lim_{t\to \infty}\frac{1}{t}\int^{t}_{0}u(s)ds=\lim_{\lambda\to 0}\lambda \mathcal{U}(\lambda),$$
where
$$ \mathcal{U}(\lambda):=\int_0^\infty e^{-\lambda t} u(t)dt:= \bar{u}(\lambda).$$
}

{\em Proof.}
Denote
$$\displaystyle\frac{1}{t}\int^{t}_{0}u(t)dt := v(t),\quad \bar{u}(\sigma)
:=
\displaystyle\int^{\infty}_{0}e^{-\sigma t}u(t)dt.$$
Then
$$\bar{v}(\lambda)
=
\displaystyle\int^{\infty}_{\lambda}\frac{\bar{u}(\sigma)}{\sigma}d\sigma$$
by the properties of the Laplace transform.

Assume that the limit
$v(\infty) := v_{\infty}$ exists:
\begin{equation}\label{eq:138} \lim_{t
\rightarrow \infty}v(t) = v_{\infty}. \end{equation} Then,
$$v_{\infty} =
\displaystyle\lim_{\lambda \rightarrow
0}\lambda\displaystyle\int^{\infty}_{0}e^{-\lambda t}v(t)dt =
\displaystyle\lim_{\lambda \rightarrow 0}\lambda\bar{v}(\lambda).$$
Indeed
$\lambda\displaystyle\int^{\infty}_{0}e^{-\lambda t}dt = 1$, so
$$\displaystyle\lim_{\lambda \rightarrow
0}\lambda\displaystyle\int^{\infty}_{0}e^{-\lambda t}(v(t) - v_{\infty})dt
= 0,$$ and (\ref{eq:138}) is verified.

One has
\begin{equation}\label{eq:139}
\lim_{\lambda \rightarrow 0}\lambda\bar{v}(\lambda) = \lim_{\lambda
\rightarrow
0}\int^{\infty}_{\lambda}\frac{\lambda}{\sigma}\bar{u}(\sigma)d\sigma =
\lim_{\lambda \rightarrow 0}\lambda\bar{u}(\lambda), \end{equation}
as follows from a simple calculation:
\begin{equation}\label{eq:140} \lim_{\lambda \rightarrow
0}\int^{\infty}_{\lambda}\frac{\lambda}{\sigma}\bar{u}(\sigma)d\sigma =
\lim_{\lambda \rightarrow
0}\int^{\infty}_{\lambda}\frac{\lambda}{\sigma^2}\sigma\bar{u}(\sigma)d\sigma
= \lim_{\sigma \rightarrow 0}\sigma\bar{u}(\sigma), \end{equation} where
we have used the relation
$\displaystyle\int^{\infty}_{\lambda}\displaystyle\frac{\lambda}{\sigma^2}d\sigma
= 1$.$\\$ Alternatively, let $\sigma^{-1} = \gamma$.
Then, \begin{equation}\label{eq:141}
\int^{\infty}_{\lambda}\frac{\lambda}{\sigma^2}\sigma\bar{u}(\sigma)d\sigma
=
\frac{1}{1/\lambda}\int^{1/\lambda}_{0}\frac{1}{\gamma}\bar{u}(\frac{1}{\gamma})d\gamma
=
\frac{1}{\omega}\int^{\omega}_{0}\frac{1}{\gamma}\bar{u}(\frac{1}{\gamma})d\gamma.
\end{equation} If $\lambda \rightarrow 0$, then $\omega =
\lambda^{-1} \rightarrow \infty,$ and if
$$\psi :=\gamma^{-1}\bar{u}(\gamma^{-1}),$$
then
\begin{equation}\label{eq:142} \lim_{\omega \rightarrow
\infty}\frac{1}{\omega}\int^{\omega}_{0}\psi d\gamma = \psi(\infty) =
\lim_{\gamma \rightarrow 0}\gamma^{-1}\bar{u}(\gamma^{-1})
= \lim_{\sigma \rightarrow 0}\sigma\bar{u}(\sigma). \end{equation}
Lemma 1 is proved. \hfill$\Box$
$$ $$
 \textbf{Lemma 2} {\em Equation (\ref{eq:17}) holds.}

 {\em Proof.} As $a \rightarrow 0$, one has
\begin{equation}\label{eq:235} \frac{\partial}{\partial
N_s}\frac{e^{-\sqrt{\lambda}|s - s'|}}{4\pi|s - s'|} =
\frac{\partial}{\partial N_s}\frac{1}{4\pi|s - s'|} +
\frac{\partial}{\partial N_s} \frac{e^{-\sqrt{\lambda}|s - s'|} -
1}{4\pi|s - s'|}. \end{equation} It is known (see \cite{R476}) that
\begin{equation}\label{eq:236}
\int_{\mathcal{S}_m}ds\int_{\mathcal{S}_m}\frac{\partial}{\partial
N_s}\frac{1}{4\pi|s - s'|}\sigma_m(s')ds' = -\frac{1}{2}
\int_{\mathcal{S}_m}\sigma_m(s')ds' = -\frac{1}{2} Q_m. \end{equation}
On the other hand, as $a \rightarrow 0$, one has
\begin{equation}\label{eq:237} \bigg|
\int_{\mathcal{S}_m}ds\int_{\mathcal{S}_m}\frac{e^{-\sqrt{\lambda}|s -
s'|} - 1}{4\pi|s - s'|}\sigma_m(s')ds' \bigg| \leq
|Q_m|\int_{\mathcal{S}_m}ds\frac{1-e^{-\sqrt{\lambda}|s - s'|}}{4\pi|s
- s'|} = o(Q_m). \end{equation}
The relations (\ref{eq:236}) and
(\ref{eq:237}) justify (\ref{eq:17}).

 Lemma 2 is proved. \hfill$\Box$
$$ $$

\textbf{Lemma 3} {\em If assumption (\ref{eq:14}) holds, then inequality (\ref{eq:13})
holds.}

 {\em Proof.} One has \begin{equation}\label{eq:332} \mathcal{J}_{1, m} :=
|g(x, x_m)Q| =  \frac{|Q_m|e^{-\sqrt{\lambda}|x - x_m|}}{4\pi|x - x_m|},
\end{equation}
and
\begin{equation}\label{eq:333} \mathcal{J}_{2, m} \le
\frac{e^{-\sqrt{\lambda}|x - x_m|}}{4\pi|x -
x_m|}\max\bigg( \sqrt{\lambda}a, \frac{a}{|x - x_m|}
\bigg)\int_{\mathcal{S}_m}|\sigma_m(s')|ds'
\end{equation}
where $|x - x_m| \geq d$, and $d>0$ is the smallest distance between two neighboring
particles. One may consider only those values of $\lambda$ for which
$\lambda^{1/4} a<\frac{a}{d}$, because for the large values of $\lambda$,
such that $\lambda^{1/4}\ge \frac {1}{d}$
the value of  $e^{-\sqrt{\lambda}|x - x_m|}$ is negligibly small.
The average temperature depends on the behavior of $\mathcal{U}$
for small $\lambda$, see Lemma 1.

One has $|Q_m|=\int_{\mathcal{S}_m}|\sigma_m(s')|ds'>0$
 because $\sigma_m$ keeps sign on $\mathcal{S}_m$, as follows
from equation (\ref{eq:20}) as $a \rightarrow 0$.$\\$
It follows from
(\ref{eq:332}) - (\ref{eq:333}) that \begin{equation}\label{eq:334} \bigg|
\frac{\mathcal{J}_{2,m}}{\mathcal{J}_{1,m}} \bigg| \leq O\bigg( \bigg|
\frac {a}{x - x_m} \bigg| \bigg) \leq O\bigg( \frac{a}{d} \bigg) << 1.
\end{equation} From (\ref{eq:334}) by the arguments similar to the given
in \cite{R509} one obtains (\ref{eq:13}).

Lemma 3 is proved.  \hfill$\Box$

\end{document}